\documentclass[]{aa}  

\def\arcsec{\hbox{$^{\prime\prime}$}}

\usepackage{graphicx}
\usepackage{txfonts}
\usepackage[colorlinks=false]{hyperref}
\begin{document}


\title{Magnetic field fluctuations in the shocked umbral chromosphere}

   \author{T. Felipe,
          \inst{1,2}\fnmsep\thanks{tobias@iac.es}
          S. J. Gonz\'alez Manrique,
          \inst{1,2,3,4}
            C. R. Sangeetha,
          \inst{1,2}
          \and
          A. Asensio Ramos\inst{1,2}
          }

   \institute{Instituto de Astrof\'{\i}sica de Canarias 
              38205 C/ V\'{\i}a L{\'a}ctea, s/n, La Laguna, Tenerife, Spain
         \and
             Departamento de Astrof\'{\i}sica, Universidad de La Laguna
             38205, La Laguna, Tenerife, Spain 
         \and
             Leibniz-Institut für Sonnenphysik (KIS), Schöneckstr. 6, 79104 Freiburg, Germany
         \and  
             Astronomical Institute, Slovak Academy of Sciences, 05960 Tatranská Lomnica, Slovak Republic\\
             }

   \date{Received ; accepted }

 \titlerunning{Magnetic field Fluctuations in \ion{He}{I} 10830 \AA{}}
  \authorrunning{Felipe et al.}
  
 
  \abstract
   {Umbral chromospheric observations show the presence of magnetoacoustic shocks. Several recent studies have reported magnetic field fluctuations associated with those shock waves. The mechanism behind these periodic magnetic field changes is still an unsolved question. }
   {We aim to study the properties and origin of magnetic field fluctuations in the umbral chromosphere.}
   {Temporal series of spectropolarimetric observations were acquired with the GREGOR telescope on 2017 June 18. The chromospheric and photospheric conditions, including the temporal evolution of the magnetic field, were derived from simultaneous inversions of the \ion{He}{I}~10830~\AA{} triplet and the \ion{Si}{I}~10827~\AA{} line using HAZEL2 code. The oscillations are interpreted using wavelet analysis and context information from UV observations acquired with the Atmospheric Imaging Assembly on board the Solar Dynamics Observatory (SDO/AIA) and the Interface Region Imaging Spectrograph (IRIS).}
   {The chromospheric magnetic field shows strong fluctuations in the sunspot umbra, with peak field strengths up to 2900 G. These inferred field strength is comparable to the magnetic field strength in the upper photosphere. Magnetic field and velocity umbral oscillations exhibit a strong coherence, with the magnetic field lagging the shock fronts detected in the velocity fluctuations. This points to a common origin of the fluctuations in both parameters, whereas the analysis of the phase shift between photospheric and chromospheric velocity is consistent with upwards wave propagation. These results suggest that the strong inferred magnetic field fluctuations are caused by changes in the response height of the \ion{He}{I}~10830~\AA{} line to the magnetic field, which is sensitive to high photospheric layers after the shock fronts. The analysis of EUV data shows a weak brightening in a coronal loop rooted in the umbra around the time of the measured magnetic field fluctuations. This coronal activity could possibly have some impact on the inferred fluctuations, but it is not the main driver of the magnetic field oscillations since they are found before the EUV event takes place.}
   {Chromospheric magnetic field fluctuations measured with the \ion{He}{I}~10830~\AA{} triplet arise due to variations in the opacity of the line. After strong shocks produced by the propagation of slow magnetoacoustic waves, the response of the line to the magnetic field can be shifted down to the upper photosphere. This is seen as remarkably large fluctuations in the line of sight magnetic field strength.}

   \keywords{Solar chromosphere --- Sunspots --- Solar atmosphere --- Solar oscillations}

   \maketitle
%

\section{Introduction}\label{sec:intro}

Magnetohydrodynamic waves are a fundamental constituent of the solar magnetized atmosphere \citep{1982ApJ...253..386L,1997SoPh..172...69H,2003SoPh..218...85B,2006ApJ...640.1153C,2006ApJ...647L..77M,2006ApJ...648L.151J,2010ApJ...722..131F,2011ApJ...735...65F,2013SoPh..287..107R,2015LRSP...12....6K,2017ApJ...847....5K,2019ApJ...871..155R,2020A&A...638A...6S}. These waves arise when acoustic p-modes interact with the surface magnetic field \citep{1993ApJ...402..721C} or can be generated by in-situ magnetoconvection taking place in active regions \citep{2008ApJ...684L..51J,2017ApJ...836...18C}. They can then propagate to the higher solar atmosphere through magnetic field elements such as sunspots, plage, or magnetic bright points \citep{1977A&A....55..239B}. The detection and study of the properties of these waves play an important role in understanding the dynamics of the solar atmosphere and their contribution to the heating of the outer layers. Most of the studies so far have focused on Doppler and/or intensity fluctuations since they are relatively easy to measure and many of the wave modes leave an imprint on these parameters. 

Over the past years, observations have also revealed oscillations in the sunspot magnetic field. Photospheric analyses have reported magnetic fluctuations with various properties \citep{1997SoPh..172...69H,1998A&A...335L..97R,1998ApJ...497..464L,1999SoPh..187..389B,1999ApJ...518L.123N,2000ApJ...534..989B,2000SoPh..192..403N,2003ApJ...588..606K,2003A&A...403..297M,2012SoPh..279..295L,2020A&A...635A..64G,2021RSPTA.37900175N, 2021A&A...654A..50N}, including a broad range in their amplitude (between a few Gauss and $\sim$300~G) and period (from a few minutes to several days). The origin of these oscillations is still not well understood due to the lack of consistency among different works. The situation is even more complex in the chromosphere, where the assumption of local thermodynamic equilibrium (LTE) is not valid, and magnetic field inferences rely on sophisticated and computationally expensive inversions under non-LTE \citep{2015A&A...577A...7S, 2019A&A...623A..74D}.

Despite these challenges, in recent years some works have addressed the study of chromospheric magnetic field variations associated with the shocks that arise due to the steepening of magnetoacoustic waves in sunspots. These shocks usually manifest as brightenings in the core of some chromospheric lines, known as umbral flashes \citep{1969SoPh....7..351B}. Independent analyses of the same spectral line have revealed contradictory results. The first attempt to measure magnetic field fluctuations in the \ion{Ca}{II} 8542~\AA{} line found no indications of oscillations in the sunspot umbra, but detected oscillations with an amplitude of $\sim$200 G in the penumbra \citep{2013A&A...556A.115D}. \citet{2017ApJ...845..102H} reported a weaker magnetic field during umbral flashes, whereas \citet{2018A&A...619A..63J} found that the magnetic field can be up to $\sim$270 G stronger in umbral flashes. However, the analysis of synthetic \ion{Ca}{II} 8542~\AA{} profiles, constructed by synthesizing this line in simulated atmospheres, revealed that the degeneracy of the inversion problem can lead to the inference of spurious magnetic field fluctuations that are not present in the actual atmospheric model \citep{2021ApJ...918...47F}, suggesting caution with the interpretation of magnetic field fluctuations inferred from the \ion{Ca}{II} 8542~\AA{} line.  

In this study, we aim to investigate chromospheric magnetic field oscillations using the \ion{He}{I} 10830~\AA{} line. This line has been previously employed to analyze magnetic field fluctuations by \citet{2018ApJ...860...28H}, who reported changes in the transversal magnetic field with amplitude up to $\sim$200 G. The organization of the paper is as follows: in Sect. ~\ref{sec:data} we describe the observations and the analysis methods, in Sect.~\ref{sec:results} we present the results, in Sect.~\ref{sec:discussion} we discuss our findings and, finally, results are summarised in Sect.~\ref{sec:conclusions}.

\section{Observations and Data Analysis} \label{sec:data}

\subsection{GREGOR/GRIS observations}

\begin{figure}[ht]  
\centering
\includegraphics[width=0.40\textwidth]{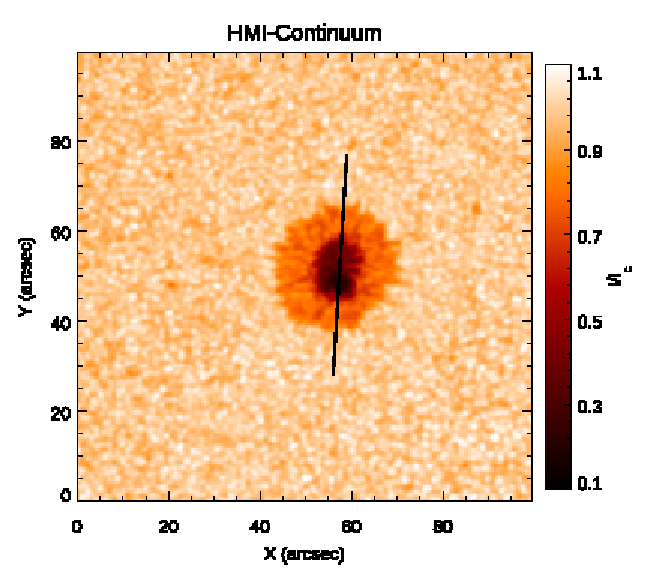}
\caption{HMI continuum intensity map of the sunspot NOAA 12662 on 2017 June 09:15 UT. The approximate position of the GRIS slit is shown by the black line in the HMI continuum intensity map.}\label{icplots}
\end{figure}

We observed the sunspot NOAA 12662 on 2017 June 18 (located near the disk center, $\mu=0.97$, with $\mu$ defined as the cosine of the heliocentric angle) using the GREGOR Infrared Spectrograph \citep[GRIS,][]{2012AN....333..872C} attached to the GREGOR telescope \citep{2012AN....333..796S}. The spectral window covers a region of approximately 18 \AA{}, including the \ion{He}{I} 10830 \AA{} line triplet and the neighboring \ion{Si}{I} 10827 \AA{} line, among other spectral lines. The full Stokes spectra were acquired with an exposure time of $100$ ms and ten accumulations. Standard polarimetric calibration \citep{1999ASPC..184....3C,2003SPIE.4843...55C} was applied to the observations, with the calibration data obtained from the GREGOR polarimetric calibration unit \citep{2012AN....333..854H}. Standard dark and flat-field reductions were applied to the data. Data were acquired between 08:02 and 09:37 UT by placing the spectrograph slit at a fixed position crossing the middle of the sunspot. The analysis presented in this work is restricted to a temporal series of roughly 37 min (starting at 08:02 UT) with a temporal cadence of 5.6 s (a total of 393 Stokes spectra). These data have been previously analyzed in \citet{2020ApJ...900L..29F} and \citet{2021NatAs...5....2F}.

Figure~\ref{icplots} shows the continuum intensity taken from the space-borne telescope Helioseismic and Magnetic Imager \citep[HMI;][]{2012SoPh..275..207S} on-board Solar Dynamics Observatory \citep[SDO;][]{Pesnell+etal2012}, including the approximate location of the spectrograph's slit. In this study, we will focus on the analysis of the sunspot umbra.

\subsection{Spectropolarimetric inversions}
We have used the HAnle and ZEeman Light v2.0 {\footnote{The HAZEL2 code can be found in \texttt{https://github.com/aasensio/hazel2}. The
user manual with detailed instructions for the usage of the code and precautions to be taken can be found in \texttt{https://aasensio.github.io/hazel2}.}} \citep[HAZEL2;][]{2008ApJ...683..542A} code to invert the \ion{He}{I} triplet, along with the \ion{Si}{I} and telluric lines. The simultaneous inversion of these two neighboring lines is carried out to account for their wings due to their very close proximity to the \ion{He}{I} triplet. HAZEL2 incorporates the effect of atomic level polarization and Paschen-Back, Hanle and Zeeman effects for the \ion{He}{I} triplet. The simultaneous inversion of the \ion{Si}{I} line is computed with the Stokes Inversion based on Response functions  \citep[SIR;][]{1992ApJ...398..375R} code, whereas the telluric line is fitted with a Voigt profile. Hence, the output from HAZEL2 inversions provides the atmospheric information from both the photosphere (derived from the \ion{Si}{I} 10827 \AA{}) and the chromosphere (derived from the \ion{He}{I} 10830 \AA{}).

The inversion is performed through an iterative process where an initial guess atmosphere is perturbed until the output radiation from the model matches the observed Stokes profiles. In our \ion{Si}{I} 10827 \AA{} inversions, we have used the hot sunspot model from \citet{1994A&A...291..622C} as the initial guess atmosphere. The above-mentioned perturbations are applied at some selected optical depths, known as nodes. The process can be repeated several times (so-called cycles) with a different number of nodes each, where the initial guess atmosphere from a cycle is given by the output from the previous cycle. Table \ref{tab:nodes} shows the inversion scheme employed in our analysis, that is, the number of nodes selected for each of the three cycles. In the case of the \ion{He}{I} triplet, HAZEL2 inversions employ a cloud model where the atmospheric parameters are constant in a slab above the solar surface. Hence, in Table \ref{tab:nodes} a 0 indicates that the corresponding atmospheric parameter is not inverted, while 1 means it is inverted.

\begin{figure}[ht]  
\centering
\includegraphics[width=0.45\textwidth]{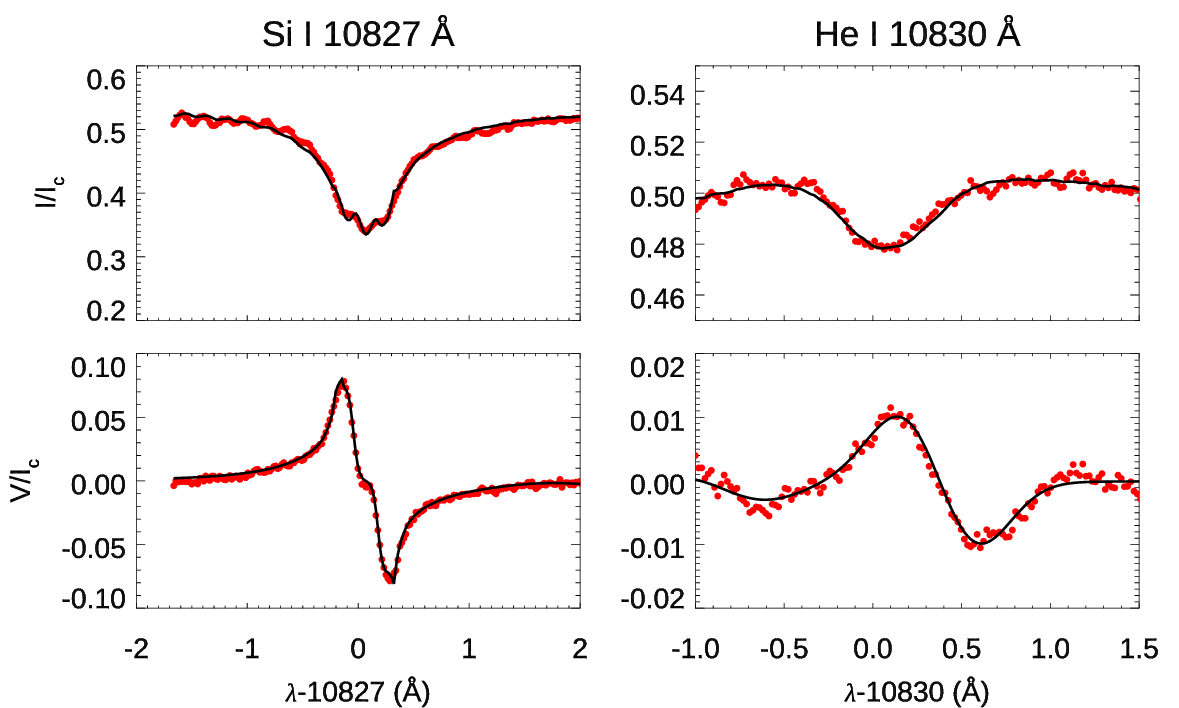}
\caption{Observed (red dots) and inverted (black lines) Stokes I (top panel) and V (bottom panel) profiles at a randomly chosen time and umbral location for \ion{Si}{I} 10827~\AA{} (left panels) and \ion{He}{I} 10830~\AA{} (right panels). }\label{stokes}
\end{figure}

The Stokes profiles were fitted using a two-component model for each pixel. One of the components corresponds to the inferred atmosphere whereas the other accounts for the stray light. The stray light represents a spurious light coming from distant spatial locations that contaminates the signal measured at a specific spatial position. This is due to the varying seeing conditions, diffraction effects, and the optical properties of the telescope. We employed a constant stray-light profile for all locations and times in the temporal series. It was computed as a spatio-temporal average of a quiet Sun region around the observed sunspot. The inversions assume that the radiation observed from a resolution element is produced by the joint contribution of the solar atmosphere at that position and the spurious stray light. The amount of stray light is variable (depends on the location and time step) and is given by the filling factor, a free parameter that is also inverted. Figure ~\ref{stokes} shows a comparison between the observed and inverted profiles at a randomly chosen time and umbral location. The inversion provides a good fit of the observed Stokes I and V profiles for both spectral lines. Independent inversions were also carried out without accounting for the stray-light correction. They exhibit significant quantitative differences (for example, in the magnetic field strength), but the qualitative results are similar.

\begin{table}[ht]
\centering
\caption{\footnotesize Inversion scheme for the photosphere (\ion{Si}{I} 10827 \AA{}) and chromosphere (\ion{He}{I} 10830 \AA{}). The table defines the number of nodes selected for temperature (T), magnetic field ($B_{x}$, $B_{y}$, and $B_{z}$), line of sight (LOS) velocity (v), microturbulence (v$_{mic}$), enhancement factor ($\beta$), and optical depth of the slab ($\tau$). }
\begin{tabular}{cccc}
\hline
{\bf Parameter} & \multicolumn{3}{c} {\bf Cycle}  \\[0.5ex]
     & {\bf 1} & {\bf 2} & {\bf 3}   \\[0.5ex]
\hline
\multicolumn{4}{c} {\bf Photosphere \ion{Si}{I} 10827} \\[0.5ex]
\hline
 T  & 3  & 4 & 5  \\[0.5ex]
v  & 1  & 2 & 4 \\[0.5ex]
$B_{x}$  & 1  & 2 & 3  \\[0.5ex]
$B_{y}$  & 1  & 2 & 3  \\[0.5ex]
$B_{z}$  & 1  & 2 & 3  \\[0.5ex]
v$_{mic}$  & 1  & 2 & 2 \\[0.5ex]
\hline
\multicolumn{4}{c} {\bf Chromosphere \ion{He}{I} 10830}  \\[0.5ex]
\hline
v  & 1  & 1 & 1 \\[0.5ex]
$B_{x}$  & 0  & 1 & 1  \\[0.5ex]
$B_{y}$  & 0  & 1 & 1  \\[0.5ex]
$B_{z}$  & 0  & 1 & 1  \\[0.5ex]
$\beta$ & 0  & 1 & 1  \\[0.5ex]
$\tau$ & 1  & 1 & 1 \\[0.5ex]
\hline
\end{tabular}
\label{tab:nodes}
\end{table}

\subsection{SDO/AIA observations}
The interpretation of the results is supported by data from the Atmospheric Imaging Assembly \citep[AIA][]{Lemen+etal2012} instrument onboard SDO. AIA acquires observations of the full solar disk in seven extreme ultraviolet (EUV) bands (94 \AA, 131 \AA, 171 \AA, 193\AA, 211 \AA, 304 \AA, and 335 \AA) with a temporal cadence of 12 s, in addition to two ultraviolet wavelengths (1600 and 1700 \AA) with a cadence of 24 s. The EUV wavelengths effectively respond to temperatures from 10$^{5.5}$ to 10$^{7.5}$ K \citep{O'Dwyer+etal2010}, whereas the UV emission is dominated by contributions from the lower solar atmosphere. The data has a spatial scale of 0.6\arcsec pixel$^{-1}$. We use level 1.0 data co-temporal to the GREGOR/GRIS observations.

\subsection{IRIS observations}

We have also examined observations from the Interface Region Imaging Spectrograph \citep[IRIS;][]{dePontieu+etal2014} as context images of the chromosphere and transition region. Between 07:55:56 and 08:14:02 UT (partially overlapping our GREGOR observations), IRIS was acquiring a raster map of the same active region. We are interested in the slit-jaw images. They were taken in filters in the near-ultraviolet (NUV; 2830 \AA\ and 2796 \AA) and far-ultraviolet (FUV; 1330 \AA\ and 1400 \AA) with a temporal cadence of 68 s, for a total of 16 images per filter. These slit-jaw filters mostly sample the upper photosphere (2830 \AA), chromosphere (2796 \AA), upper chromosphere (1330 \AA), and low transition region (1400 \AA).

\section{Results} \label{sec:results}

\subsection{Chromospheric velocity and magnetic field fluctuations}

\begin{figure*}[ht]  
\centering
\includegraphics[width=0.85\textwidth]{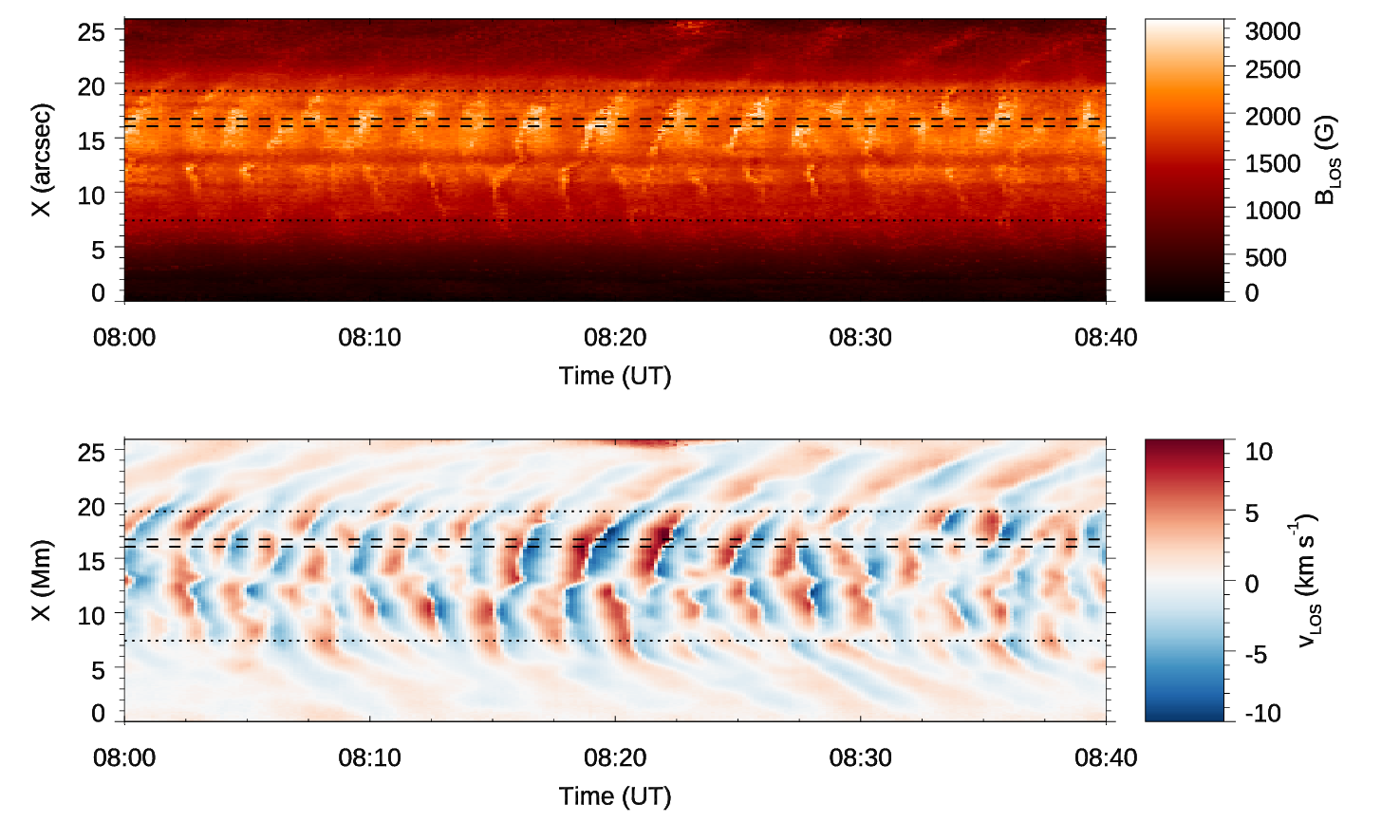}
\caption{Temporal evolution of the chromospheric magnetic field (top panel) and LOS velocity (bottom panel) along the spectrograph slit, as inferred from the inversions of \ion{He}{I} triplet. Horizontal dotted lines indicate the umbra-penumbra boundaries. Horizontal dashed lines mark the spatial locations illustrated in Figs. \ref{B_V_temporal_Udark} and \ref{B_V_temporal_Udots}.}\label{B_V_inv}
\end{figure*}

Figure~\ref{B_V_inv} illustrates the inverted maps of chromospheric LOS magnetic field and velocity derived from HAZEL2. As expected, stronger magnetic fields are found in the umbra, especially in the umbral region seen darker in continuum intensity (Fig. \ref{icplots}), where the LOS magnetic field periodically fluctuates in the range 1900-2900 G. Other umbral locations exhibit a weaker magnetic field, with a typical strength around 1600 G. 

The velocity signal shows the well-known umbral chromospheric oscillations in the three-minute band. Clear indications of shocks are also found, with sudden changes from red (downflows) to blue (upflows). In the penumbral regions, the pattern of running penumbral waves is clearly seen as wavefronts that reach farther radial distances as time increases. Interestingly, the phase of the wavefronts exhibits coherence in the dark umbra (and towards the right-hand side penumbra), but a marked change in the phase is found at $x\sim 29-30$\arcsec, the approximate location of the innermost umbral dot. This may indicate some differences in the excitation of the waves in both regions \citep[magnetoconvection taking place in umbral dots and light bridges has been suggested as the driver of the three-minute chromospheric oscillations][]{2017ApJ...836...18C} or differences in the travel time of the waves along the magnetic field lines.

\begin{figure*}[ht]  
\centering
\includegraphics[width=0.85\textwidth]{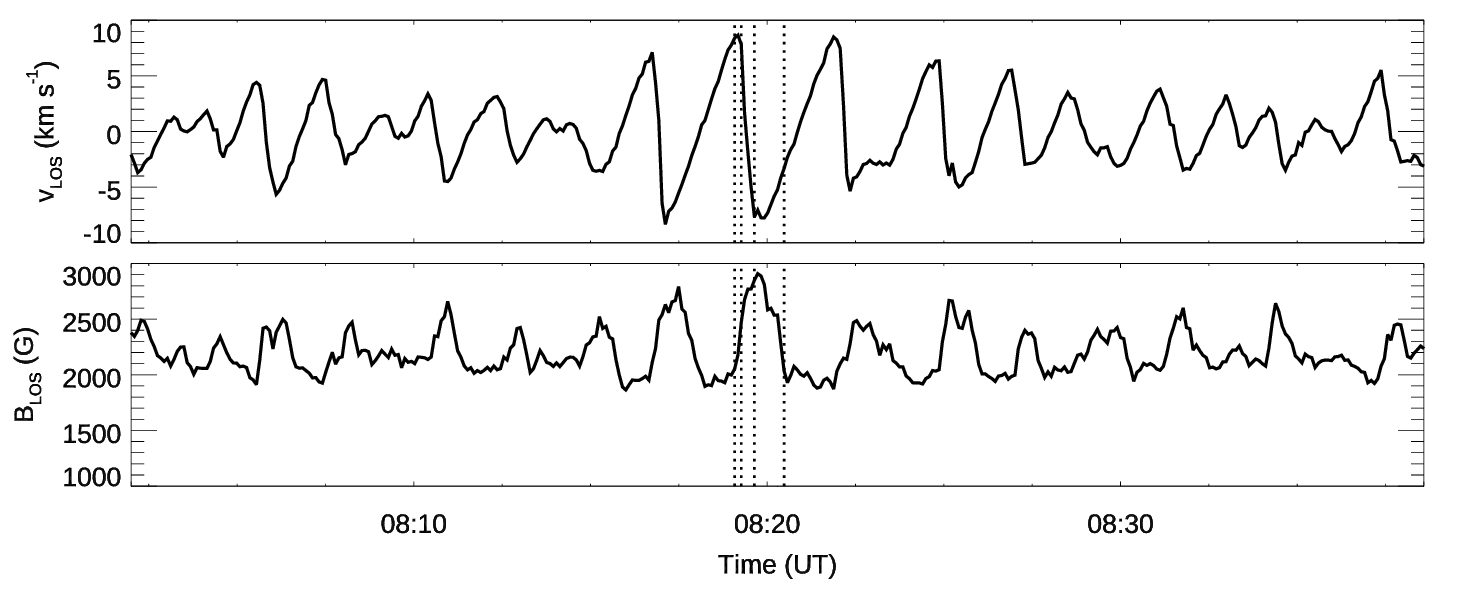}
\caption{Temporal evolution of LOS velocity (top panel) and magnetic field (bottom panel) fluctuations inferred from the inversions of the \ion{He}{I} 10830 \AA\ line in the umbral location indicated by the bottom dashed line from Fig. \ref{B_V_inv}. Magnetic field fluctuations have been smoothed by averaging in three time steps windows. Vertical dotted lines indicate the temporal steps illustrated in Fig. \ref{fig:stokes_evolution}.}\label{B_V_temporal_Udark}
\end{figure*}

\begin{figure*}[ht]  
\centering
\includegraphics[width=0.85\textwidth]{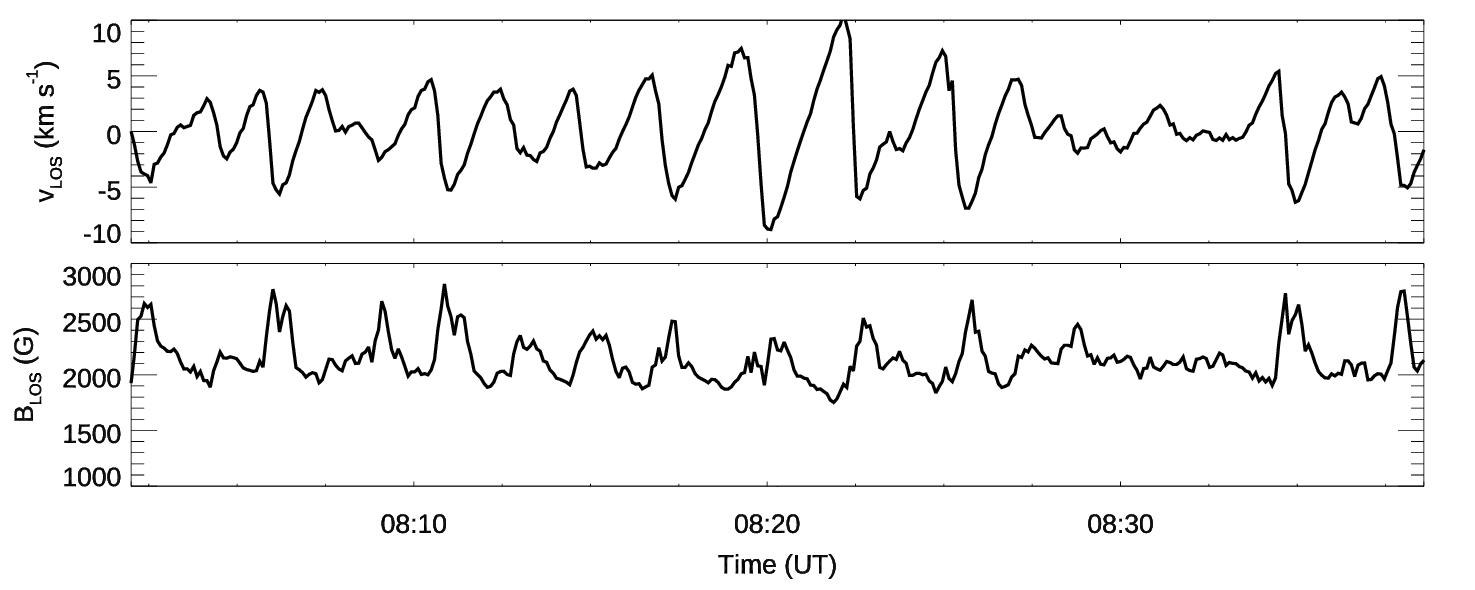}
\caption{Same as Figure \ref{B_V_temporal_Udots} but for the location indicated by the top dashed line from Fig. \ref{B_V_inv}.}\label{B_V_temporal_Udots}
\end{figure*}

Figures \ref{B_V_temporal_Udark} and \ref{B_V_temporal_Udots} show the temporal evolution of the LOS magnetic field and velocity at two different locations of the sunspot umbra. The maximum amplitude of velocity oscillations is around 10~kms$^{-1}$. This high amplitude, comparable to the local sound speed, leads to the development of chromospheric shocks, which are seen as a progressive rise in the velocity evolution followed by a steeper fall after reaching the maximum amplitude. These sudden velocity changes, characteristic of the shock, are generally accompanied by peaks in the LOS magnetic field. In the umbra, the background magnetic field is around 2000~G. Magnetic field enhancements up to 900~G above that background field strength are found to lag the velocity signal. The stronger magnetic field excursions found in Fig. \ref{B_V_temporal_Udark} take place after the stronger velocity shocks (around 08:20 UT), but smaller peaks with more modest magnetic field strength increments of $\sim$300-500~G are measured during the whole temporal series. Strong magnetic field excursions are not always associated with the highest velocity amplitudes. For example, at around 08:06 UT in Fig. \ref{B_V_temporal_Udots} a magnetic field peak of 2700 G is found during a velocity shock with 5~kms$^{-1}$ amplitude.

{

\subsection{\ion{He}{I} 10830 \AA\ spectral profiles during magnetic field excursions}

Figure \ref{fig:stokes_evolution} illustrates the temporal evolution of \ion{He}{I} 10830 \AA\ Stokes I and V profiles during the development of one of the largest magnetic field fluctuations in the sunspot umbra. The first illustrated time step is just before a shock. The line exhibits a large Doppler shift to the red (v$_{\rm LOS}$=8.39 km s$^{-1}$), but the magnetic field is still at the quiescent stage, with a strength around 1900 G. The second and third time steps (two middle rows) capture the development of the shock (see Fig. \ref{B_V_temporal_Udark}). Between these times the Doppler velocity changes from 7.82 km s$^{-1}$ to -7.67 km s$^{-1}$ over a temporal span of 22.4 s, while the LOS magnetic field strength shows a striking enhancement. During this shock, the absorption depth of the line is greatly reduced, but no emission is found in \ion{He}{I} 10830 \AA\ intensity. At the last time step, the line is again in the quiescent state, where the probed magnetic field strength has returned to the background values in the range 1900-2000 G.

\begin{figure}[ht]  
\centering
\includegraphics[width=0.45\textwidth]{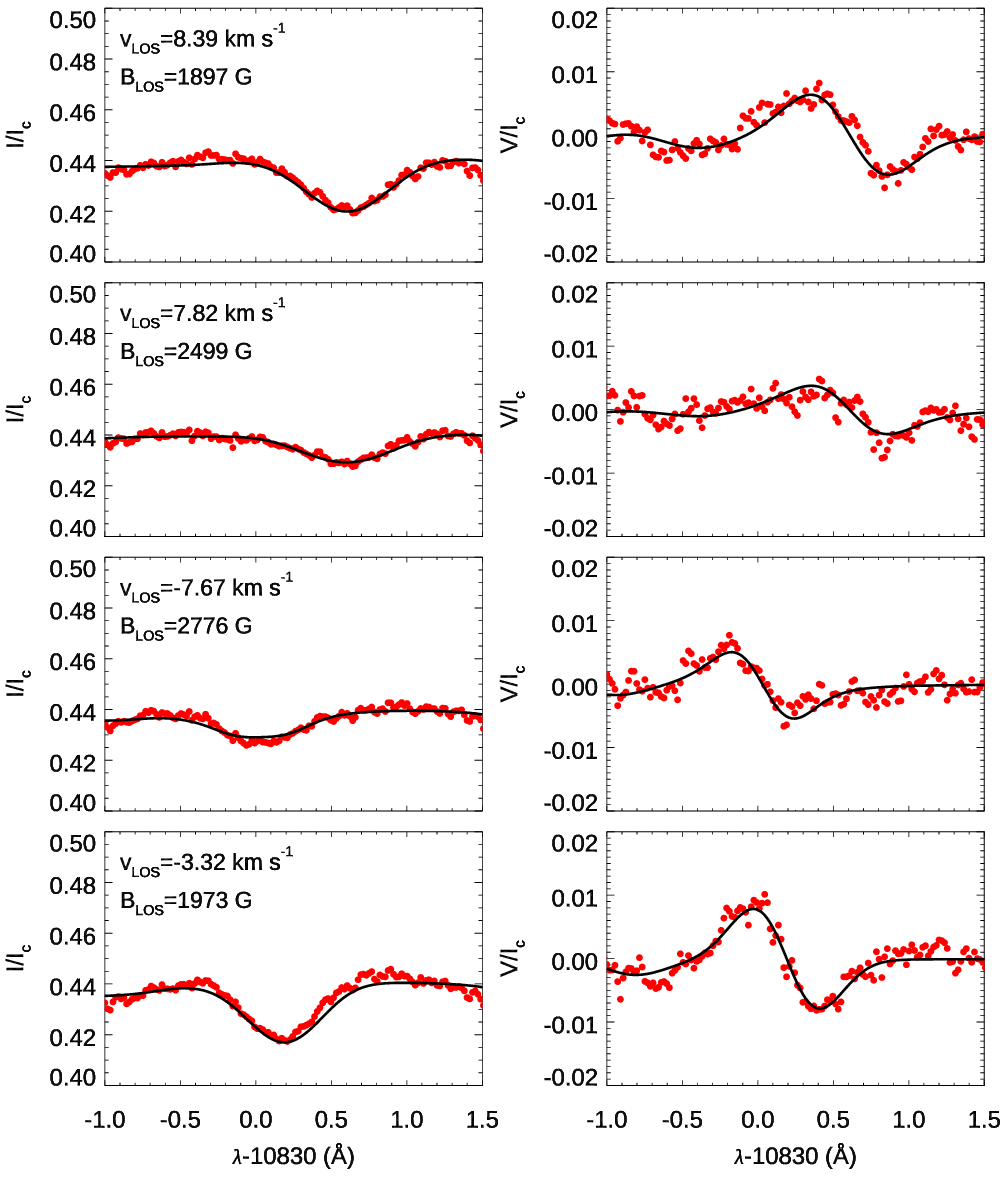}
\caption{Observed (red dots) and inverted (black lines) Stokes I (left panels) and V (right panels) profiles of the \ion{He}{I} 10830~\AA{} line during the development of a large amplitude magnetic field fluctuation. Each row corresponds to a different time step (indicated by vertical dotted lines in Fig. \ref{B_V_temporal_Udark}), with time increasing from top to bottom. }\label{fig:stokes_evolution}
\end{figure}

\subsection{UV data analysis}

The magnetic field strength measured during the peaks is much higher than that inferred with the \ion{He}{I}~10830~\AA{} triplet in other sunspots \citep[e.g.,][]{Joshi+etal2017}. In contrast, strong magnetic fields have been measured with the \ion{He}{I}~10830~\AA{} \citep{2021ApJ...916....5S} and the \ion{He}{I} D$_{\rm 3}$ \citep{Libbrecht+etal2019} lines after energetic events taking place at higher coronal layers. To extend our study to the transition region and corona and support the interpretation of the results, we have employed EUV images acquired with SDO/AIA and NUV and FUV data from IRIS.

A visual inspection of the EUV data prior to and during the GREGOR temporal series employed in this study revealed a brightening in one of the coronal loops rooted to the sunspot umbra. Figure \ref{fig:AIA_mapas} shows maps of SDO/AIA data at three selected filters and several times during this brightening. Coronal loops (especially visible in 171 \AA) are only present on one side of the sunspot. The brightening, visible in the three filters but more clearly in 304 \AA, develops in the loop marked by a blue-dotted line during those time steps. The temporal evolution of the EUV intensity along this loop is illustrated in Fig. \ref{fig:AIA_timedistance}. At around 08:10 UT, some brightness increase is detected at $s=20\arcsec$, where $s$ is the distance measured along the loop. The intensity peak is reached 4-5 min later. During all the event, the brightening is restricted to the same region around $s=20\arcsec$. The temporal evolution of the average signal in the brightening region (white square in Fig. \ref{fig:AIA_av}) is shown in Fig. \ref{fig:AIA_av}. All the filters exhibit a progressive increase in intensity between 08:10 and 08:13 UT, followed by a sudden intensity enhancement. Later, the EUV emission in the brightening region returns to approximately the same values it has before the event. The beginning of this brightening is also visible in the IRIS slit-jaws images at 2796 \AA, 1330 \AA, and 1400 \AA. However, we do not discuss it in depth since the finish time of IRIS observations (08:14 UT) before the peak of the event does not allow a comprehensive study.

At the sunspot umbra, no significant changes in the EUV intensity are found. We also do not find enhanced emission in IRIS slit-jaws at 1330 \AA\ and 1400 \AA. Only the 2796 \AA\ filter often exhibits clear emission from umbral flashes \citep{Tian+etal2014}. The examination of the umbra in NUV, FUV, and EUV shows no indications that the inferred magnetic field fluctuations are associated with energetic events taking place at the transition region or corona.

\begin{figure*}[ht]  
\centering
\includegraphics[width=0.85\textwidth]{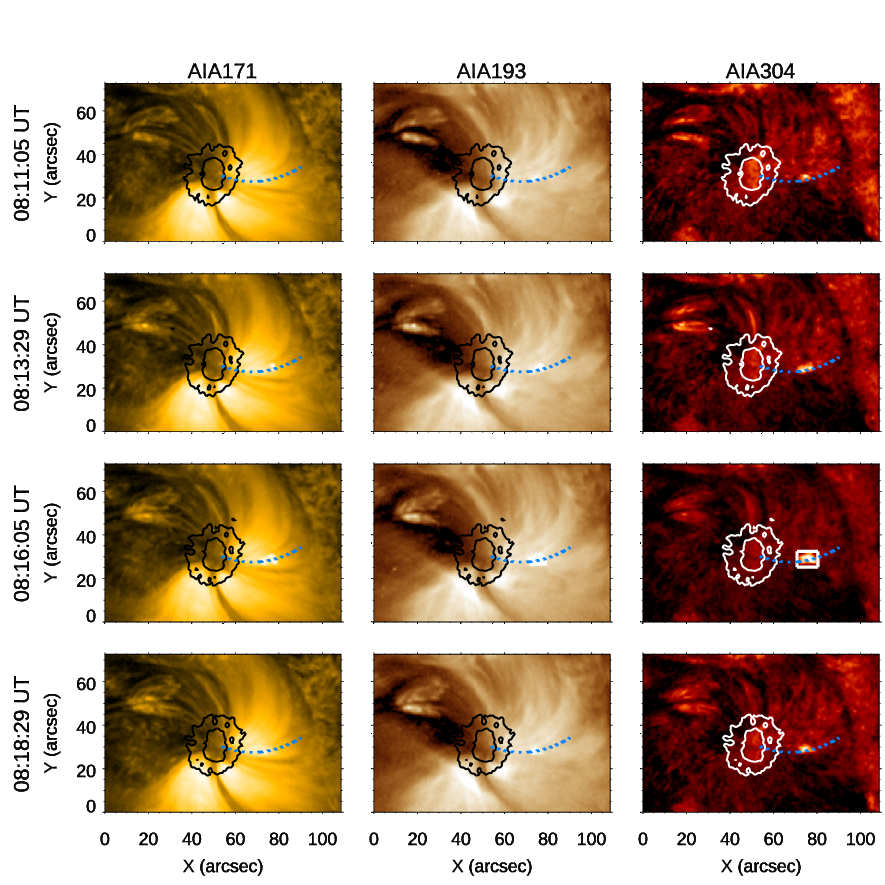}
\caption{EUV observations of the sunspot. Each column corresponds to a different AIA filter (from left to right: 171 \AA, 193 \AA, and 304 \AA). Rows represent different times, with the time indicated at the left-hand side. Black/white lines illustrate the umbra and penumbra boundaries as selected from contours of constant intensity in AIA 1700 \AA. The blue-dotted line traces the coronal loop illustrated in Fig. \ref{fig:AIA_timedistance}. The white square delimits the region used for the average plotted in Fig. \ref{fig:AIA_av}     }\label{fig:AIA_mapas}
\end{figure*}

\begin{figure*}[ht]  
\centering
\includegraphics[width=0.85\textwidth]{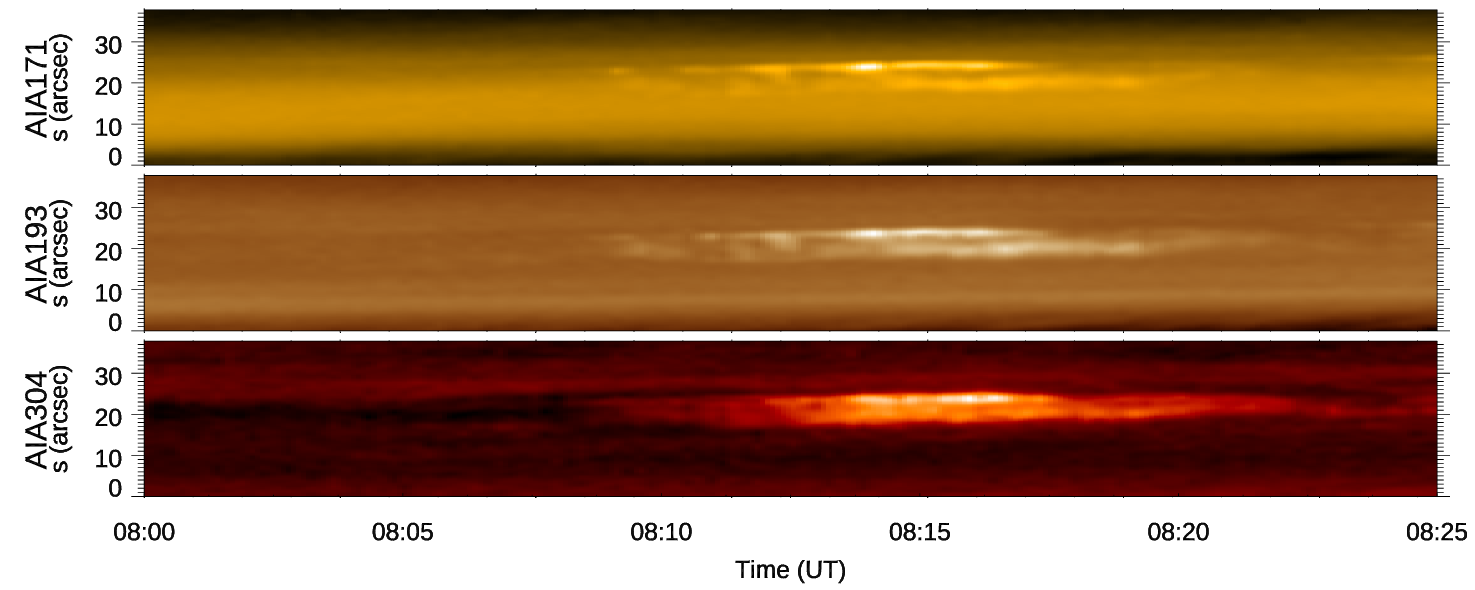}
\caption{EUV intensity along a coronal loop. Temporal evolution of the intensity in AIA filters at 171 \AA\ (top panel), 193 \AA\ (middle panel), and 304 \AA\ (bottom panel) at all positions along the blue-dotted lane represented in Fig. \ref{fig:AIA_mapas}. s=0 correspond to the footpoint of the coronal loop at the sunspot umbra.} \label{fig:AIA_timedistance}
\end{figure*}

\begin{figure}[ht]  
\centering
\includegraphics[width=0.45\textwidth]{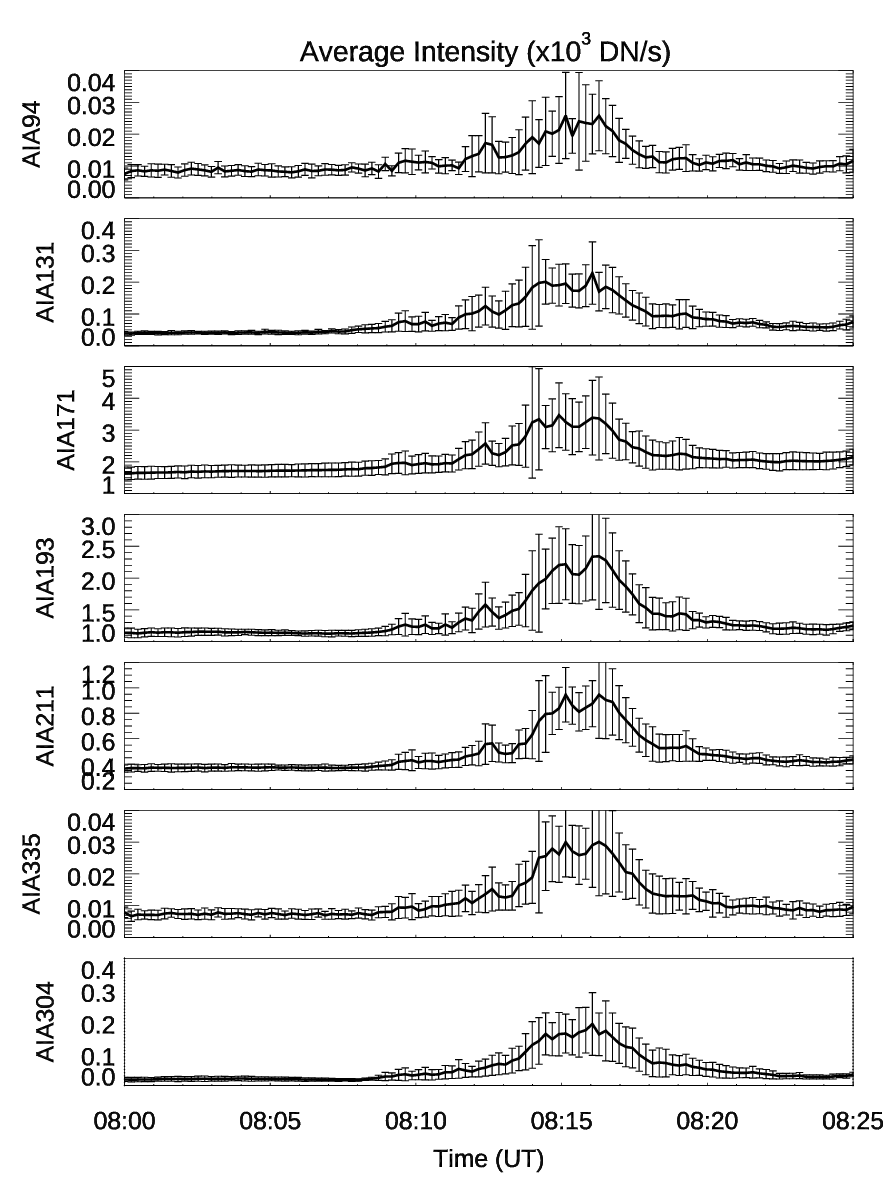}
\caption{Temporal evolution of the average EUV intensity in the brightening region (white square in Fig. \ref{fig:AIA_mapas}). Error bars show the standard deviation. The AIA filters are indicated at the left-hand side of each panel.} \label{fig:AIA_av}
\end{figure}

\subsection{Wavelet analysis}

To further explore the oscillatory nature of the velocity and magnetic field fluctuations, we have performed a wavelet analysis \citep{Torrence+Compo1998} of both variables. The wavelet transform (WT) decomposes a time series into time and frequency domains, allowing the determination of the dominant modes and their temporal evolution. Wavelet analysis can also be employed to compute the wavelet cross spectrum (WCS) between two time series. This quantity exhibits a large value when both signals have large power at similar frequencies and around the same time and, more interesting, it can be used to derive the phase difference as a function of time and frequency. Wavelet analysis is a common approach for the study of oscillations in the solar atmosphere \citep[e.g.,][]{Bloomfield+etal2004,Lohner-Bottcher+Bello-Gonzalez2015,Guevara-Gomez+etal2021}.

Figure \ref{fig:wavelet_power} shows the wavelet power of the photospheric and chromospheric velocities and the power of the WCS between both signals for one umbral location. The photospheric and chromospheric velocity power are consistent with the results from many previous studies of umbral oscillations. Photospheric velocity oscillations are dominated by fluctuations in the 5-minute band (3-4 mHz), whereas at the chromosphere the main power is shifted to the 3-minute band (with a peak at around 6-7 mHz in this dataset). The power also exhibits some variations during the temporal series. As expected, higher chromospheric power is found during the time steps where the amplitude of the chromospheric oscillations is higher. Interestingly, at those times (08:15-08:25 UT) there is also a photospheric power increase in the 3-minute band (around 6 mHz). This photospheric power is possibly the predecessor for the chromospheric counterpart \citep{2006ApJ...640.1153C}. 

The phase angle of the WCS provides the phase difference between the two signals employed for its computation. This is equivalent to computing the phase of the Fourier cross spectra, but the use of wavelets allows the evaluation of the changes in the phase difference with time. In the following, only the information from spectral regions inside the cone of influence (discarding those parts where the edge effects are relevant) and with a confidence level in the WCS above 95\% are considered for the interpretation of the results. We have computed the phase angle for all the locations belonging to the dark part of the umbra (20 spatial positions, excluding the umbral region where umbral dots are abundant). Figures \ref{fig:wavelet_dphase_pre} and \ref{fig:wavelet_dphase_post} illustrate the measured phase difference and coherence (as a function of frequency) between the photospheric and chromospheric velocities for two selected times. Each circle in the top panels represents the phase difference at one umbral position at a given frequency. A positive phase shift indicates that the chromospheric velocity lags the photospheric velocity. The bottom panels show the coherence spectra. They indicate the statistical significance of the phase difference between two signals, according to the phases measured for a certain number of pairs of signals (in our case, 20 pairs). In addition to the confidence level of the WCS, we also employ coherence as an index to validate the phase results. We consider the measured phase differences to be relevant when the coherence is above 0.7 (horizontal black dashed lines in bottom panels from Figs. \ref{fig:wavelet_dphase_pre} and \ref{fig:wavelet_dphase_post}).

Figure \ref{fig:wavelet_dphase_pre} shows the phase difference spectra between photospheric and chromospheric velocities at 08:13 UT, that is, just before the sudden brightening in the EUV emission (Fig. \ref{fig:AIA_av}) and before a strong vertical magnetic field is inferred from the \ion{He}{I}~10830~\AA{} triplet (Fig. \ref{B_V_temporal_Udark}). The phase shift of frequencies below $\sim$3.5 mHz is not reliable since both the confidence and coherence are low. Above that frequency, the phase shift progressively increases, which is a clear indication of the presence of upward wave propagation between the two layers probed by the \ion{Si}{I} 10827~\AA{} and \ion{He}{I}~10830~\AA{} lines. 

After the EUV brightening ($\sim$ 08:20 UT), the phase spectrum between photospheric and chromospheric velocities exhibits a similar behavior (Fig. \ref{fig:wavelet_dphase_post}), that is, the phase difference increases with the frequency. However, at this time step, the slope of the increment is lower. This indicates a lower phase shift between the velocity in both layers and points to some differences in the probed oscillations. To illustrate these differences, we have fitted the phase shift to a model of linear wave propagation in a gravitationally stratified isothermal atmosphere with radiative losses, following \citet{2006ApJ...640.1153C}. The model has three parameters: temperature ($T_{\rm 0}$), height difference between both signals ($\Delta z$), and the radiative cooling time ($\tau_{\rm R}$). For the temperature, we selected the average umbral temperature inferred from the inversion of the \ion{Si}{I} 10827~\AA{} line ($T_{\rm 0}=4160$ K). The other two parameters were obtained from a manual fit of the phase difference to the model. We found that the phase spectrum prior to the peak of the EUV brightening is better explained by a model with $\Delta z=900$ km, in close agreement with \citet{2006ApJ...640.1153C} and \citet{2010ApJ...722..131F}. In contrast, a few minutes after the EUV brightening, the phase spectrum is best fitted with a model with $\Delta z=450$ km.

Figure \ref{fig:wavelet_VB} illustrates the wavelet analysis of the WCS and the phase spectra between the LOS velocity and LOS magnetic field measured at the chromosphere from the inversion of the \ion{He}{I}~10830~\AA{} line. A highly significant (above the 95\% confidence level and total coherence) phase delay of 133$^{\circ}\pm 28^{\circ}$ is measured, with the magnetic field lagging the velocity. This is consistent with the velocity and magnetic field fluctuations illustrated in Figs. \ref{B_V_temporal_Udark} and \ref{B_V_temporal_Udots}.

\begin{figure}[ht] 
\centering
\includegraphics[width=0.45\textwidth]{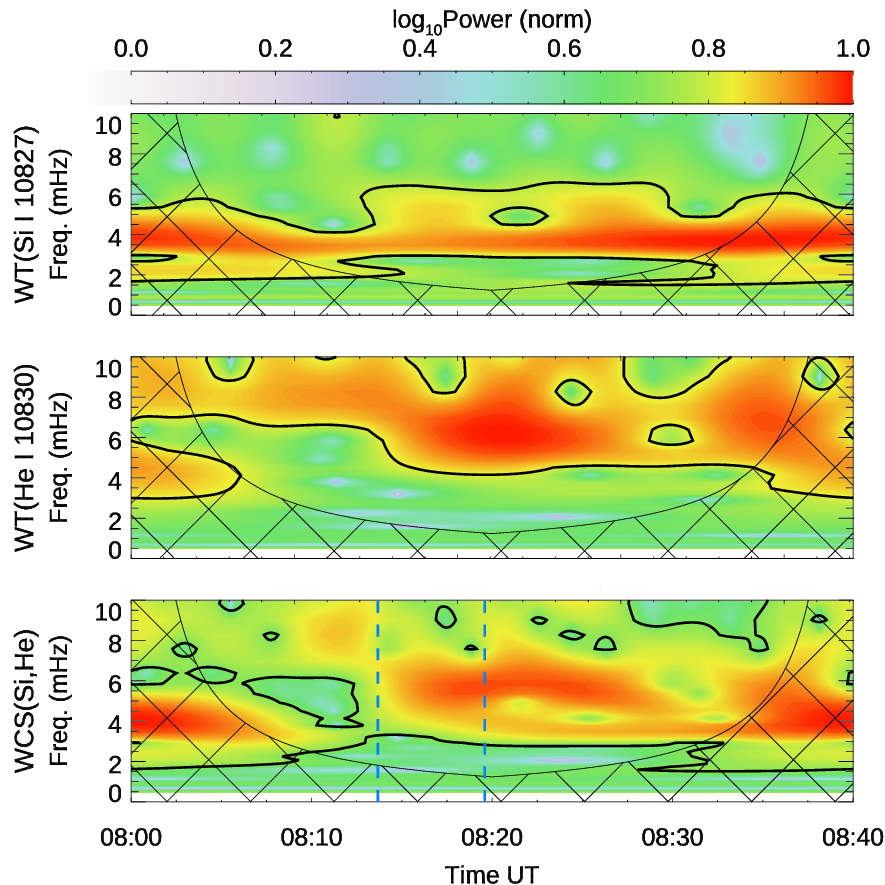}
\caption{Wavelet power of the velocity at a randomly chosen umbral location. Top panel: wavelet power of the photospheric velocity derived from the \ion{Si}{I}~10827~\AA{} line. Middle panel: wavelet power of the chromospheric velocity derived from the \ion{He}{I}~10830~\AA{} line. Bottom panel: Power of the wavelet cross spectrum between the photospheric and chromospheric velocities. The gridded region indicates the parts of the spectra out of the cone of influence. Black solid lines mark the 95\% confidence level. Vertical blue-dashed lines in the bottom panel indicate the times illustrated in Figs. \ref{fig:wavelet_dphase_pre} and \ref{fig:wavelet_dphase_post}.} \label{fig:wavelet_power}
\end{figure}

\begin{figure}[ht] 
\centering
\includegraphics[width=0.45\textwidth]{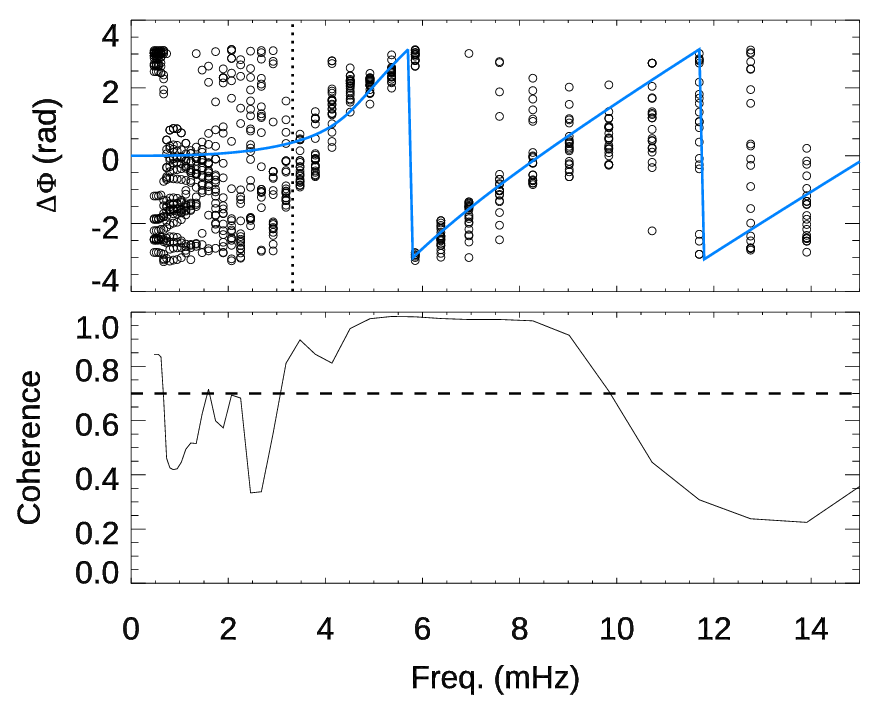}
\caption{Wavelet phase difference between the photospheric and chromospheric umbral velocities around 08:13 UT. Top panel: Phase difference at every location in the selected region. The power of the WCS of frequencies higher than that indicated by the vertical dotted line is above the 95\% confidence level. The blue line represents the expected phase difference for linear wave propagation in a gravitationally stratified atmosphere with radiative losses ($T_{\rm 0}=4160$ K, $\Delta z$=900 km, $\tau_{\rm R}$=20). Bottom panel: Coherence of the phase difference. The phase difference where the confidence is below 95\% (frequencies below the vertical dotted line in the top panel) and with coherence below the threshold (horizontal dashed line in the bottom panel) are unreliable. 
} \label{fig:wavelet_dphase_pre}
\end{figure}

\begin{figure}[ht] 
\centering
\includegraphics[width=0.45\textwidth]{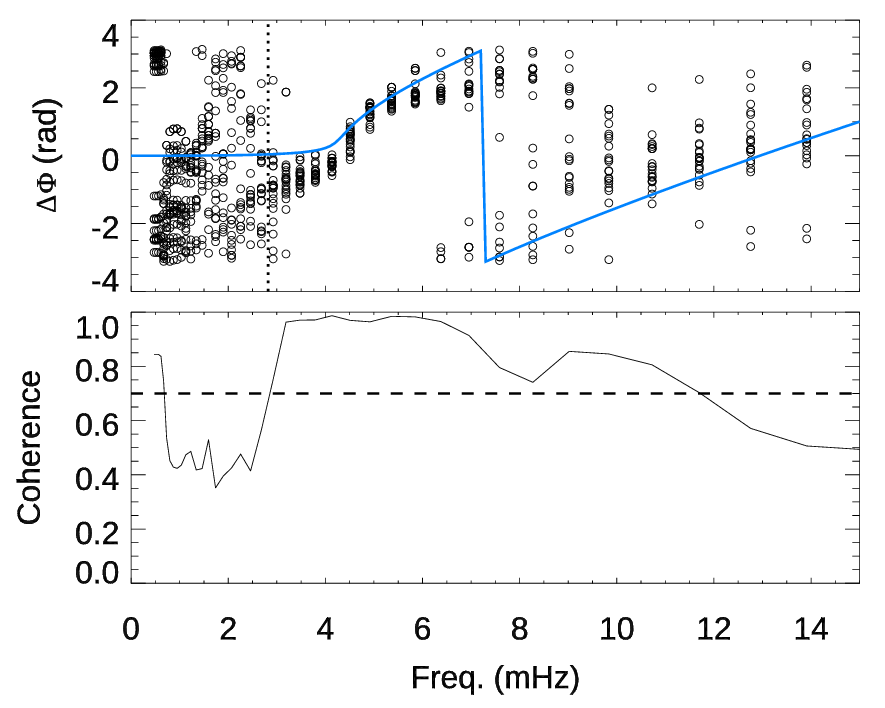}
\caption{Same as Fig. \ref{fig:wavelet_dphase_pre} but at around 08:20 UT. The parameters of the theoretical model illustrated by the blue line are $T_{\rm 0}=4160$ K, $\Delta z$=450 km, and $\tau_{\rm R}$=10.} \label{fig:wavelet_dphase_post}
\end{figure}

\begin{figure}[ht] 
\centering
\includegraphics[width=0.45\textwidth]{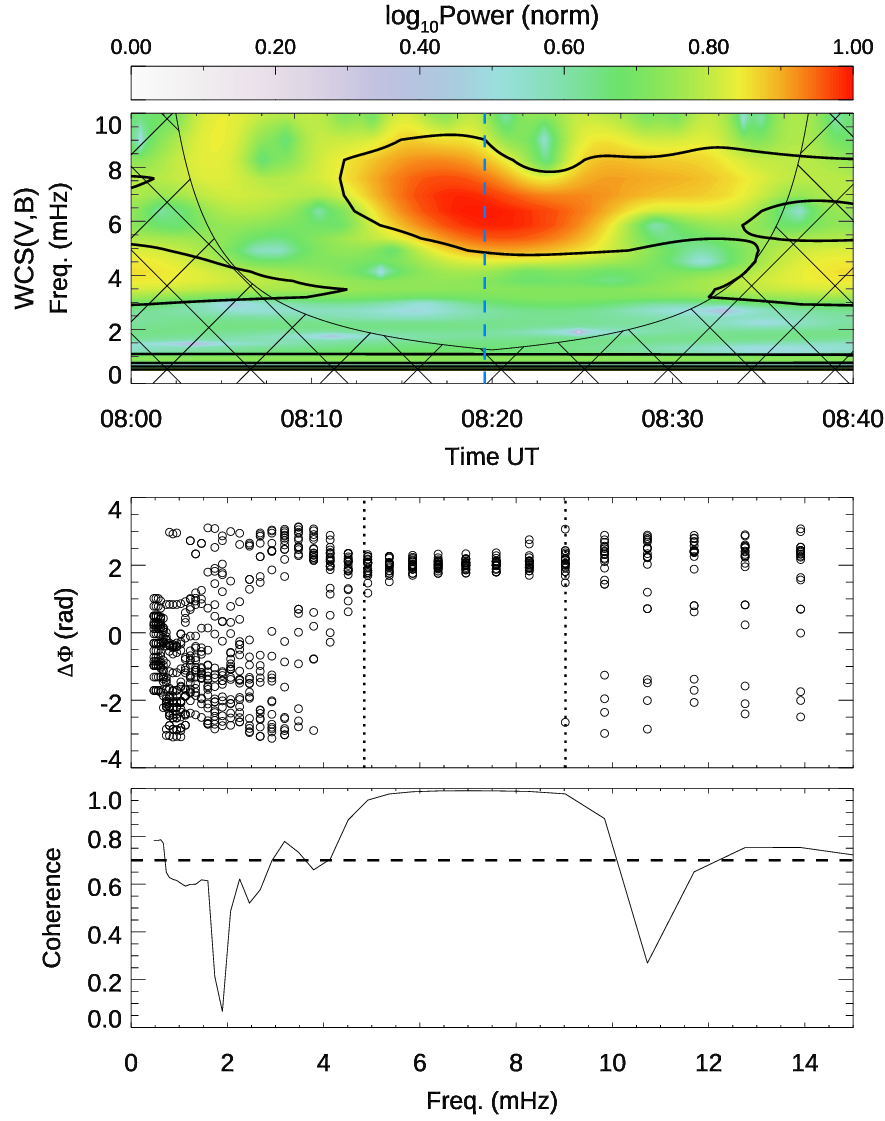}
\caption{Wavelet analysis of velocity and magnetic field fluctuations measured with the \ion{He}{I}~10830~\AA{} triplet. Top panel: Power of the wavelet cross spectrum between the velocity and magnetic field fluctuations. The gridded region indicates the parts of the spectra out of the cone of influence. Black solid lines mark the 95\% confidence level. The vertical blue-dashed line indicates the time illustrated in the spectra from middle and bottom panels. Middle panel: Wavelet phase difference between the velocity and magnetic field fluctuations in the umbra at 08:20 UT. Vertical dotted lines delimit the region with the power of the WCS above the 95\% confidence level. Bottom panel: Coherence of the phase difference. The horizontal dashed lines indicates the 0.7 threshold. } \label{fig:wavelet_VB}
\end{figure}

\section{Discussion}\label{sec:discussion}

\subsection{Magnetic field fluctuations and opacity effect}

The interpretation of the large LOS magnetic field fluctuations measured in our observations poses a compelling challenge. In the umbra, the LOS magnetic field increases from $\sim$2000 G to more than 2900 G (Fig. \ref{B_V_temporal_Udark}), possibly the strongest magnetic field strength ever measured with the \ion{He}{I} 10830 \AA\ line. Other independent inversions of the \ion{He}{I} 10830 \AA\ line in sunspots (analyzing raster maps instead of temporal series with short scanning cadence) have inferred magnetic field strengths around 1500 G \citep{Joshi+etal2017,Lindner+etal2023}. This strong magnetic field is hard to be understood as a manifestation of the actual chromospheric magnetic field. Instead, we hypothesize that these large fluctuations in the magnetic field are produced by remarkable changes in the response height of the \ion{He}{I} 10830 \AA\ triplet to the magnetic field.

We have evaluated the vertical magnetic field gradient in our observations by comparing the photospheric and chromospheric magnetic fields. The photospheric magnetic field at $\log \tau = -2$ is inferred from the \ion{Si}{I} 10827 \AA\ line, whereas the chromospheric field is derived from the \ion{He}{I} 10830~\AA{} triplet. Figure ~\ref{opaeff} shows the variation of the magnetic field at both atmospheric layers as a function of the position along the slit. For each spatial position, the median of the complete temporal series has been computed. As expected, the magnetic field strength decreases with height. The umbral magnetic field strength in the photosphere is higher by a factor of 1.1-1.4 compared to the chromosphere. This factor is slightly lower than that measured by \citet{Joshi+etal2017}, but consistent since our estimation of the photospheric magnetic field corresponds to a higher photospheric layer.

Figure \ref{fig:Bampl_Bgrad} illustrates the correlation between the amplitude of the \ion{He}{I} 10830 \AA\ magnetic field fluctuations ($\delta B_{\rm LOS}$[ch]) and the magnetic field gradient between the photosphere and chromosphere ($\Delta B_{\rm LOS}$[ph-ch]). In this diagram, each red dot represents a location in the umbra. $\Delta B_{\rm LOS}$[ph-ch] was computed as the difference between the two median magnetic fields shown in Fig. \ref{opaeff}. The following procedure was carried out to determine $\delta B_{\rm LOS}$[ch]: (i) the \ion{He}{I} 10830 \AA\ magnetic field temporal series at the selected location was smoothed by averaging in 7 time-steps windows ($\sim$39 seconds), (ii) the time steps of all the maximums (minimums) were estimated by selecting those steps whose field strength is higher (lower) than the two adjacent times, (iii) all the maximums (minimums) with a magnetic field strength above (below) the 90\% (10\%) percentile were averaged, and (iv) the difference between the averaged maximum and minimum was computed. Step (iii) was performed to discard local maximums (minimums) that are not representative of magnetic field peaks (valleys).

A positive correlation between magnetic field fluctuations measured in \ion{He}{I} 10830~\AA{} and the photosphere-chromosphere gradient is found at those places where $\Delta B_{\rm LOS}$[ph-ch]$\geq$ 500 G. The strongest magnetic field fluctuations measured in \ion{He}{I} 10830~\AA{} are found at those locations where the photospheric magnetic field is much higher than the chromospheric field. This result strongly supports the interpretation of the \ion{He}{I} 10830~\AA{} magnetic field oscillations as changes in the response height of the line, at least for the strongest magnetic field peaks. At umbral locations with lower $\Delta B_{\rm LOS}$[ph-ch] no clear correlation is found in Fig. \ref{fig:Bampl_Bgrad}. The dispersion in the results is expected. First, the amplitude of the measured magnetic field fluctuations will depend on the magnitude of the excursions in the response of the line and, thus, $\Delta B_{\rm LOS}$[ph-ch] does not fully characterize the amplitude of the oscillations. Second, we are evaluating the fluctuations in the magnetic field along the LOS. Variations due to changes in the orientation of the magnetic field are also expected to coexist with the opacity effect. All in all, we consider the relationship between $\delta B_{\rm LOS}$[ch] and $\Delta B_{\rm LOS}$[ph-ch] to be highly significant and a strong support of the opacity effect as the main effect behind the \ion{He}{I} 10830~\AA{} magnetic field fluctuations.

\begin{figure}[ht]  
\centering
\includegraphics[width=0.5\textwidth]{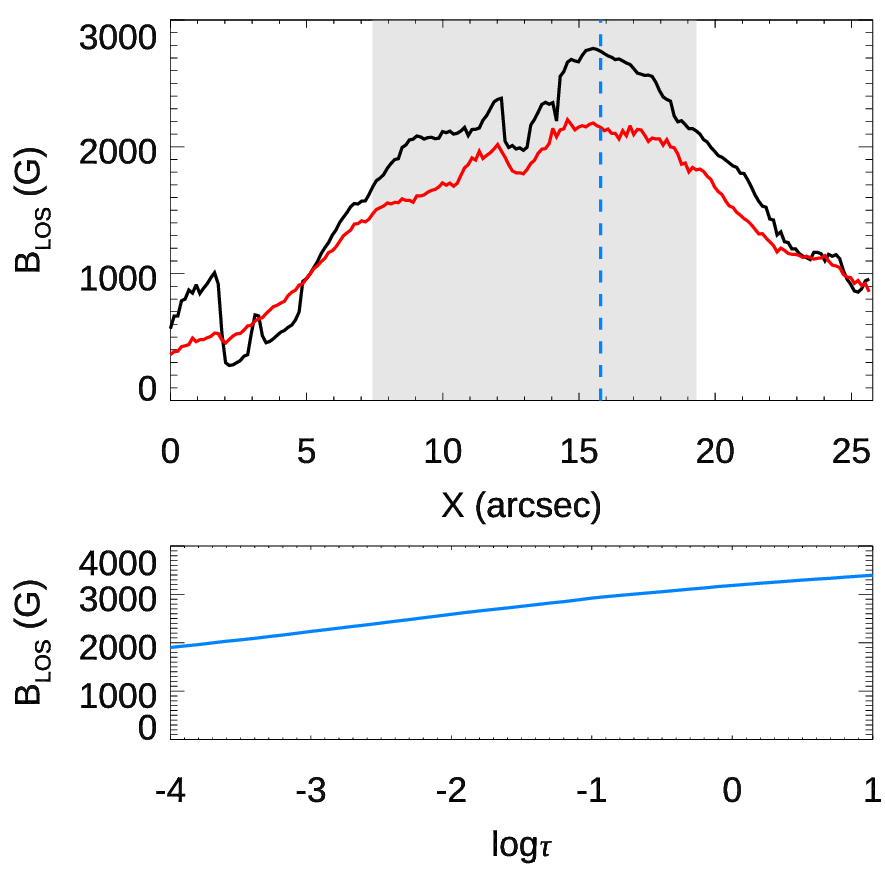}
\caption{Median of the inferred LOS magnetic field. Top panel: variation of the LOS magnetic field with the position along the slit at $\log\tau=-1.5$ (black line, derived from inversions of the \ion{Si}{I} 10827 \AA\ line) and at the chromosphere (red line, derived from \ion{He}{I} 10830 \AA\ inversions). Grey shaded area indicates the umbra region. The vertical dashed line corresponds to the location plotted in the bottom panel. Bottom panel: vertical stratification of the median photospheric magnetic field inferred with the \ion{Si}{I} 10827 \AA\ line at a selected umbral location.}\label{opaeff}
\end{figure}

\begin{figure}[ht]  
\centering
\includegraphics[width=0.5\textwidth]{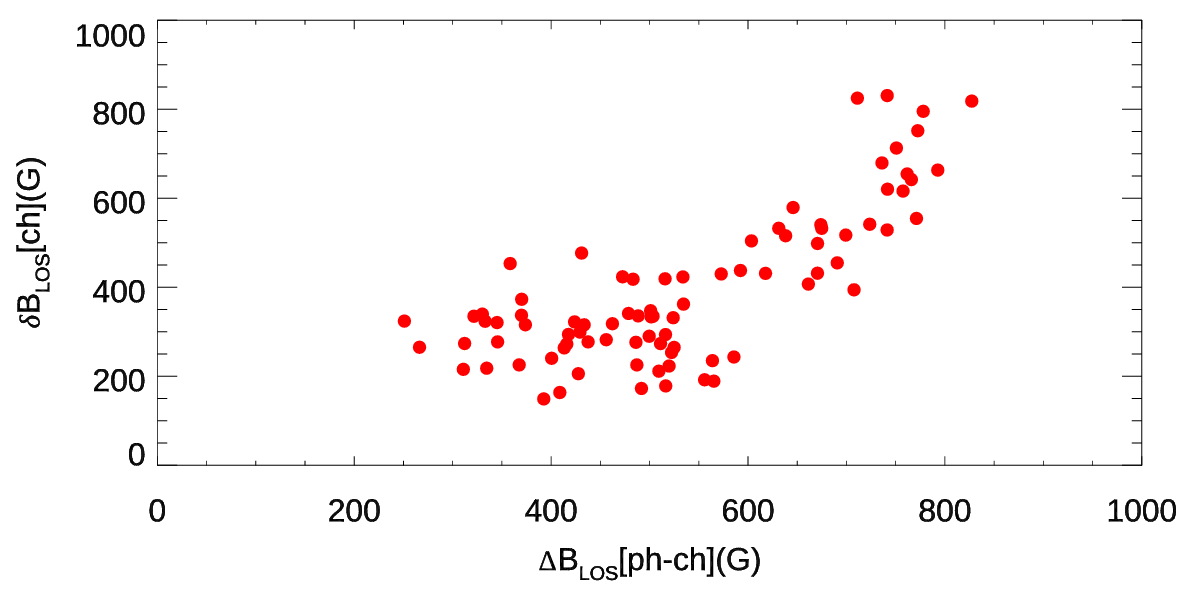}
\caption{Amplitude of the fluctuations in the LOS magnetic field plotted as a function of the variation between the photospheric and chromospheric magnetic field. Each dot represents an umbral location.}\label{fig:Bampl_Bgrad}
\end{figure}

\subsection{Origin of the magnetic field fluctuations}

The large value of the magnetic field fluctuations inferred with the \ion{He}{I} 10830~\AA{} (e.g., Fig. \ref{B_V_temporal_Udark}) and their dependence on the underneath magnetic field (Fig. \ref{fig:Bampl_Bgrad}) suggest that the \ion{He}{I} triplet probes the magnetic field of the high photosphere during shocks. In this section, we discuss the origin of those shocks that can produce such a large displacement in the height response of the line.

Previous studies have reported strong magnetic fields inferred with neutral helium in active regions associated with energetic events. \citet{2021ApJ...916....5S} reported 
LOS magnetic fields up to 2400 G in \ion{He}{I} 10830~\AA{} after a supersonic coronal downflow impacting the lower umbral atmosphere. Using observations of the \ion{He}{I} D$_{\rm 3}$, \citet{Libbrecht+etal2019} inferred a 2500 G magnetic field in a flare footpoints. \citet{Anan+etal2018} also observed a remarkable magnetic field increase in \ion{He}{I} 10830~\AA{} during a flare. All these works suggested that neutral helium lines form at deeper layers during the analyzed events.  

Here, our observations reveal similar enhancements in the magnetic field strength but also notable differences. The more relevant is that our data exhibit no trace of \ion{He}{I} 10830~\AA{} in emission. None of the large magnetic field fluctuations we have reported are associated with emission profiles. Also, we do not find enhanced emission in any of the AIA channels at the umbral locations where strong magnetic fields take place, as opposed to \citet{2021ApJ...916....5S} and \citet{Libbrecht+etal2019}. Our examination of the AIA EUV channels revealed a brightening in one of the coronal loops that is anchored to the umbra. Although this event could potentially be related to some of the measured magnetic field peaks, we note that strong magnetic fields, up to 2800 G, are also inferred before that brightening takes place (compare Figs. \ref{B_V_temporal_Udots} and \ref{fig:AIA_timedistance}).

The chromospheric umbral velocity measured in our observations is undoubtedly the signature of slow magnetoacoustic waves, as previously reported by many authors \citep[e.g.,][]{2006ApJ...640.1153C,2010ApJ...722..131F,2012ApJ...757..160J, 2015ApJ...812L..15K} and proved by the inferred upward propagation of longitudinal waves (Figs. \ref{fig:wavelet_dphase_pre} and \ref{fig:wavelet_dphase_post}). The detected magnetic field fluctuations exhibit a strong coherence with the velocity oscillations (Fig. \ref{fig:wavelet_VB}), pointing to a common origin of both signals. Thus, we conclude that the changes in the formation of the \ion{He}{I} 10830~\AA{} produced by slow magnetoacoustic wave shocks are the origin of the magnetic field enhancements.

The \ion{He}{I} 10830~\AA{} opacity is determined by the photonionisation rate in the \ion{He}{I} ground state continuum and the chromospheric electron density \citep[e.g.,][]{1975ApJ...199L..63Z, 1994IAUS..154...35A,2008ApJ...677..742C, 2016A&A...594A.104L}. The former is given by coronal and transition region irradiation, which we assume does not change significantly during the analyzed process since we do not observe intensity enhancements in AIA or IRIS observations. We speculate that fluctuations in the chromospheric electron density are the main contribution of sunspot oscillations to changes in the \ion{He}{I} 10830~\AA{} opacity, which manifests as fluctuations in the inferred magnetic field. These changes in the opacity shift the response of the line even to high photospheric layers, where the population of triplet state helium is low \citep{2008ApJ...677..742C}. The strong downflows from the shocks, combined with the longer relaxation timescales for recombination from \ion{He}{II} in a nonequilibrium state \citep{2014ApJ...784...30G}, may populate the helium triplet state in those low atmospheric layers, as suggested by \citet{2021ApJ...916....5S}.

The impact of the downflowing material coming from the brightening event identified in EUV data could potentially modify the measured oscillations. For example, we have detected some differences in the phase shift between the velocity oscillations inferred from the \ion{He}{I} 10830~\AA{} and the \ion{Si}{I} 10827 \AA\ lines (compare Figs. \ref{fig:wavelet_dphase_pre} and \ref{fig:wavelet_dphase_post}). After the brightening, the phase shift can be fitted with a theoretical model of wave propagation with a lower height difference between the atmospheric height of both signals. However, this fitting must be interpreted with care. The model assumes linear wave propagation, but recent findings indicate that umbral chromospheric oscillations are not propagating but are stationary instead \citep{Jess+etal2020, Felipe+etal2020}. The estimated height difference should be interpreted as the height difference between the formation height of the \ion{Si}{I} 10827 \AA\ line and the atmospheric layer where stationary umbral oscillations start, since above that layer the phase of the wave is constant. Also, the formation height of the lines fluctuates. These changes can be striking in the case of the \ion{He}{I} 10830~\AA{} triplet, as proven by our examination of the magnetic field fluctuations. The atmospheric layers where the triplet is sensitive change at temporal scales shorter than most of the periods probed in our wavelet analysis.

\subsection{Alternative scenarios of magnetic field fluctuations}
In this section, we discuss several models for magnetic field oscillations that can be proposed as alternatives to the opacity effect. We argue that none of them can satisfactorily explain our measurements. 

Magnetized atmospheres can support three different wave modes: fast and slow magnetoacoustic waves and the Alfv\'en mode. Strictly speaking, this separation only holds for homogeneous plasmas, where the three wave branches are effectively decoupled, but these terms are generally employed for inhomogeneous plasmas like those found in the Sun. In a medium where magnetic pressure is much higher than gas pressure, like the umbral chromosphere, slow magnetoacoustic waves behave like acoustic waves propagating mainly along the magnetic field lines. They barely produce changes in the magnetic field since their oscillations are longitudinal and, thus, their motions are mostly directed along the field lines. In contrast, fast and Alfvén waves can generate LOS magnetic field fluctuations at a given atmospheric height through two different processes \citep{1996ApJ...465..436U}: compression, where horizontal motions compress and expand the field lines density, and bending, in which the horizontal motions cause changes in the orientation of the magnetic field lines with respect to the observer. The former corresponds to fast magnetoacoustic waves, whereas the latter is associated with Alv\'en waves. The V-B phase expected for the compression mechanism is -90$^{\circ}$ (magnetic field fluctuations leading velocity fluctuations), while in the case of the bending mechanism it should be 0 \citep{1996ApJ...465..436U}. Our phase shift measurements cannot be explained by these processes (Fig. \ref{fig:wavelet_VB}). 

Under the bending mechanism scenario, we could assume that the strongest chromospheric LOS magnetic field ($\sim$2900 G) corresponds to the times when the magnetic field vector is directed along the LOS. A background magnetic field (not associated with post-shock atmospheres) of $\sim$2000 G would require the field vector to be inclined  $\sim$46$^{\circ}$ from the LOS, implying a transversal field strength around 2100 G. Such a strong transversal magnetic field would leave an imprint in the linear polarization signal, which is not present in our observations (in the umbra, no Stokes Q and U signals are detected above the noise level). In addition, \citet{2018ApJ...860...28H} found that shock waves produce small changes in the inclination of the magnetic field (below 8$^{\circ}$). Our results cannot be interpreted as the manifestation of changes in the direction of the magnetic field vector.

As previously discussed, the chromospheric velocity oscillations measured in the umbra are produced by slow magnetoacoustic waves. The strong coherence of the \ion{He}{I} 10830~\AA{} velocity oscillations with the detected magnetic field fluctuations indicates that both signals are caused by the same phenomenon. Thus, we can discard fast magnetoacoustic and Alfv\'en waves as the origin of these magnetic field oscillations. 

The propagation of waves along flux tubes imposes certain boundary conditions that lead to the existence of many oscillatory eigenmodes, which can be discriminated according to their radial structure, the number of nodes in the azimuthal direction, and their wave speed \citep{1983SoPh...88..179E}. In recent years, several works have claimed the detection of those resonant modes in sunspots \citep{2017ApJ...842...59J, 2022NatCo..13..479S}. Many detections in small magnetic flux tubes have also been reported \citep{2011ApJ...729L..18M,2013A&A...559A..88S,2017ApJS..229...10J}. Analytical works have determined the phase relations between the fluctuations produced by these wave modes in several observables \citep{2009ApJ...702.1443F, 2013A&A...551A.137M}, including the V-B phase. Both studies predict a V-B phase with either 180$^{\circ}$ or $\pm 90^{\circ}$ phase shifts, depending on the propagating/standing nature of the waves and the wave modes involved. None of these estimations can account for the 133$^{\circ}\pm 28^{\circ}$ V-B phase measured in our observations. In addition, the amplitude of the magnetic field fluctuations predicted for those eigenmodes under chromospheric conditions is significantly lower than the amplitude of the detected field fluctuations. These models were developed to support the interpretation of photospheric oscillations and do not account for some properties of the umbral chromosphere, such as the expansion of the magnetic field with height or the presence of non-linearities. Thus, the comparison with our data must be assessed with caution.

\section{Conclusions} \label{sec:conclusions}

We have reported chromospheric magnetic field fluctuations in a sunspot umbra inferred from inversions of the \ion{He}{I} 10830~\AA{} line. The large amplitude of these fluctuations, reaching LOS magnetic field strengths up to 2900 G, makes it challenging to associate them with chromospheric magnetism. Their magnetic field strength is comparable to that found in the underneath umbral photosphere. We interpret these fluctuations as the result of the opacity effect. Immediately after the shocks, the response of the \ion{He}{I} 10830~\AA{} line to the magnetic field is shifted to lower atmospheric heights. 

These magnetic field fluctuations show remarkable coherence and a well-defined phase shift with the velocity oscillations. This finding clearly indicates that they are driven by the slow magnetoacoustic waves that we detect in the velocity signal and discards the contribution from other wave modes (fast and Alfv\'en) as the origin of the magnetic field fluctuations. Also, from the examination of co-temporal UV data, we find no indications of coronal or transition region energetic events that could be impacting the lower atmosphere and driving the fluctuations.

Our observations open several questions regarding the formation of the \ion{He}{I} 10830~\AA{} line. We find that, after the shocks produced by the upward propagation of magnetoacoustic waves, the response of the \ion{He}{I} 10830~\AA{} triplet may come from atmospheric layers as low as the high photosphere. This interpretation is in agreement with the suggestion from several works that have analyzed the magnetic field inferred in neutral helium lines after flares or supersonic coronal downflows \citep{Anan+etal2018, Libbrecht+etal2019, 2021ApJ...916....5S}. Here, we found that such energetic events are not required to produce striking fluctuations in the response height of the \ion{He}{I} 10830~\AA{} triplet since they can also be produced by shocks associated with wave propagation. New observational analyses of temporal series are necessary to assess whether this is a common behavior or an uncommon occurrence due to some of the properties of the analyzed sunspot.

\begin{acknowledgements}
     Financial support from grants PGC2018-097611-A-I00, PID2021-127487NB-I00, and PGC2018-102108-B-I00, funded by MCIN/AEI/ 10.13039/501100011033 and by “ERDF A way of making Europe” is gratefully acknowledged. TF acknowledges grant RYC2020-030307-I funded by MCIN/AEI/ 10.13039/501100011033 and by “ESF Investing in your future”. SJGM is grateful for the support of the European Research Council through the grant ERC-2017-CoG771310-PI2FA, the MCIN/AEI/ 10.13039/501100011033 and “ERDF A way of making Europe” through grant PGC2018-095832-B-I00, and the project VEGA 2/0048/20. The 1.5-meter GREGOR solar telescope was built by a German consortium under the leadership of the Leibniz-Institut f\"ur Sonnenphysik in Freiburg in Freiburg with the Leibniz-Institut f\"ur Astrophysik Potsdam, the Institut f\"ur Astrophysik G\"ottingen, and the Max-Planck-Institut f\"ur Sonnensystemforschung in G\"ottingen as partners, and with contributions by the Instituto de Astrof\'isica de Canarias and the Astronomical Institute of the Academy of Sciences of the Czech Republic. The redesign of the GREGOR AO and instrument distribution optics was carried out by KIS whose technical staff is gratefully acknowledged.

\end{acknowledgements}

%
%
\bibliographystyle{aa} 
\bibliography{crs_bib.bib} 
\end{document}